\documentclass[conference]{IEEEtran}
\IEEEoverridecommandlockouts
\usepackage{cite}
\usepackage{amsmath,amssymb,amsfonts}
\usepackage{algorithmic}
\usepackage{graphicx}
\usepackage{textcomp}
\usepackage{xcolor}
\usepackage{balance}
\usepackage{listings}

\newcommand{\subf}[2]{%
  {\small\begin{tabular}[t]{@{}c@{}}
  #1\\#2
  \end{tabular}}%
}

\def\BibTeX{{\rm B\kern-.05em{\sc i\kern-.025em b}\kern-.08em
    T\kern-.1667em\lower.7ex\hbox{E}\kern-.125emX}}

\begin{document}

\title{A Fast, Scalable, Universal Approach For Distributed Data Aggregations\\

}

\author{
\IEEEauthorblockN{
Niranda Perera\IEEEauthorrefmark{1}\textsuperscript{\textsection},
Vibhatha Abeykoon\IEEEauthorrefmark{1}\IEEEauthorrefmark{2}\textsuperscript{\textsection},
Chathura Widanage\IEEEauthorrefmark{1}\textsuperscript{\textsection}, 
Supun Kamburugamuve\IEEEauthorrefmark{2}\textsuperscript{\textsection} \\
Thejaka Amila Kanewala\IEEEauthorrefmark{3},
Pulasthi Wickramasinghe\IEEEauthorrefmark{1}, \\
Ahmet Uyar\IEEEauthorrefmark{2},
Hasara Maithree\IEEEauthorrefmark{6},
Damitha Lenadora \IEEEauthorrefmark{6},
and Geoffrey Fox\IEEEauthorrefmark{1}\IEEEauthorrefmark{2}
}

\IEEEauthorblockA{\IEEEauthorrefmark{1}Luddy School of Informatics, Computing and Engineering, IN 47408, USA\\
\{vlabeyko,dnperera,pswickra\}@iu.edu}
\IEEEauthorblockA{\IEEEauthorrefmark{2}Digital Science Center, Bloomington, IN 47408, USA\\
\{cdwidana, skamburu, auyar, gcf\}@iu.edu}
\IEEEauthorblockA{\IEEEauthorrefmark{3}Indiana University Alumni, IN 47408, USA\\
thejaka.amila@gmail.com}
\IEEEauthorblockA{\IEEEauthorrefmark{6}Department of Computer Science and Engineering, University of Moratuwa, Sri Lanka\\
\{hasaramaithree.15, damitha.15\}@cse.mrt.ac.lk}
}

\maketitle

\begingroup\renewcommand\thefootnote{\textsection}
\footnotetext{These authors contributed equally.}
\endgroup

\begin{abstract}

In the current era of Big Data, data engineering has transformed into an essential field of study across many branches of science. Advancements in Artificial Intelligence (AI) have broadened the scope of data engineering and opened up new applications in both enterprise and research communities. Aggregations (also termed \emph{reduce} in functional programming) are an integral functionality in these applications. They are traditionally aimed at generating meaningful information on large data-sets, and today, they are being used for engineering more effective features for complex AI models. Aggregations are usually carried out on top of data abstractions such as tables/ arrays and are combined with other operations such as grouping of values. There are frameworks that excel in the said domains individually. But, we believe that there is an essential requirement for a data analytics tool that can universally integrate with existing frameworks, and thereby increase the productivity and efficiency of the entire data analytics pipeline. \emph{Cylon} endeavors to fulfill this void. In this paper, we present \emph{Cylon}'s fast and scalable aggregation operations implemented on top of a distributed in-memory table structure that universally integrates with existing frameworks.
\end{abstract}




\section{Introduction}

The massive amount of data generated by a wide variety of applications are used as an input to Artificial Intelligence (AI) and Machine Learning (ML) applications for further processing. However, these AI/ML applications cannot use the raw data as it is; hence the need for a preprocessing step to convert raw data into a consumable form. Aggregations are an essential class of operations used in the preprocessing stage of a data processing application. Sum, maximum, minimum, count, mean, median and standard deviation are some of the most widely used aggregation operations. They are commonly applied after categorising (grouping) data to extract the meaningful input to the AI/ML applications.

The faster we can group and aggregate data, the sooner we can get the final result from the data processing applications. The performance of the aggregation operations can be increased by executing an operation in parallel. Furthermore, with the amount of data generated today, the aggregations and grouping needs distributed execution to increase scalability. While parallel and distributed execution gives us performance and scalability, they inherently add complexity to the implementations of aggregation and grouping operations. Hiding those complexities from AI/ML users and providing them an easy-to-use, familiar abstraction is challenging, but it increases AI/ML user productivity. In this paper, we present a fast, scalable and easy-to-use "group by" and "a set of aggregation operators" implementation.

Aggregators and group by operations are implemented as part of the Cylon\cite{cylon-site}: Online Analytical Processing (OLAP) system. Cylon utilizes distributed-memory parallel execution model. In Cylon, data are distributed among multiple compute nodes which process data using Bulk Synchronous Parallel (BSP) \cite{gerbessiotis1994direct} model, employing a columnar data structure to represent data. Our implementation inherits these concepts from Cylon and contributes to the performance and scalability of the implementation. We present a framework for aggregation implementation in Cylon and we currently implement four essential distributed memory parallel aggregators using this framework: min, max, count, and sum. The framework consists of phases, each phase maximizing computation to improve the performance.
For group by, we present two approaches: Hash-based and Pipeline-based. These are discussed in detail in \ref{s:group}. In the same section we illustrate how we adopted the existing group of aggregation techniques in traditional RDBMS databases and existing Big Data frameworks for our implementation. 
The core of Cylon is developed using C/C++ to achieve maximum performance. Similar to other Cylon operators, the aggregations and group by operations implemented are also exposed with Python bindings so that they can be integrated with AI/ML applications.


For this research, our main objectives are: 1.) provide efficient compute and communication kernels for aggregation and group by operations; 2.) offer efficient language bindings; and 3.) integrate seamlessly with existing data engineering and data science systems. In Section \ref{s:data-engineering}, we discuss the role of aggregations in data engineering and high performance computing paradigms. Section \ref{s:cylon} showcases how we designed the Cylon system to facilitate high performance data engineering. Section \ref{s:aggoperators} details in depth how our aggregation operations are embedded within the Cylon system. We demonstrate a set of experiments conducted on Cylon compared to state-of-the-art data engineering systems in Section \ref{s:experiments}. Finally in Sections \ref{s:conclusion} and \ref{s:future-work}, we highlight the conclusions drawn from our work and extensions to our research respectively. 

\section{Reductions in Data Engineering}\label{s:data-engineering}



Data engineering focuses on practical applications of data collection, analysis, and prediction. It involves data extraction, transformation and loading (ETL) workloads. The \emph{transformation} phase employs relational and linear algebra operators in which data aggregation functions play a vital role. This is evident from the presence of a large number of aggregation queries in the decision support benchmarks such as TPC \cite{tpc-bench}. They also play a vital role in recent AI and ML applications.

Approaches for aggregations depend on how data is laid out on the physical memory. When looking at general applications on data engineering, data layouts can be broadly categorize into 1. tables, and 2. arrays (or tensors). In both categories, data can be laid out on row-major or column-major fashion. 

Aggregation operations are also an integral part of the grouping/ categorizing operations. Applying a aggregation (also termed \emph{reduction}) operation on grouped data, extracts a summarized the insight on the grouped data. Grouping and aggregation may reduce the size of the dataset, but the grouping operation may require substantial computational overheads, such as, moving data, randomly access memory, etc. 

\subsection{Aggregations in Tables }\label{s:s:big-data-systems}

A Table abstraction (also referred as data-frames) carry heterogeneously typed data defined by a schema.The framework architecture would choose to use row or column-major structure but aggregations/ reductions are usually carried out on a column. Hence, aggregations on table with a columnar data structure would be very efficient because it seeks contiguous memory locations and allows trivial SIMD parallelization. 

Tables are the backbone of Big Data systems. Apache Hadoop with \emph{map-reduce} \cite{apache-hadoop}\cite{dean2008mapreduce} marked the first generation of Big Data analytics. Subsequently Apache Spark\cite{apache-spark}, Apache Flink and Apache Storm were introduced, featuring better scalability and performance. These systems are designed on a JVM-based back-end. They are mostly geared for commodity cloud environments and enterprise clusters. The task-based data-flow execution in these systems promotes usability usability, but whether they achieve the native hardware performance is questionable.   

\subsection{Aggregations in Arrays/ Tensors}\label{s:s:hpc-data-engineering}

Compared to tables, arrays and tensors entail homogeneously typed multi-dimensional data. Aggregations/ reductions are carried out on a particular indices of these data. Advantages of seeking contiguous memory also holds for array data. Arrays/ tensors may have limitations in representing relational data.

These are the main data structure used for High performance computing applications in domain sciences (i.e. Physics, Chemistry, Biology, etc.). Highly optimized linear algebraic computations are available on arrays through BLAS routines. OpenMP, MPI and PGAS are some of the systems designed to provide distributed computing capability for these structures. These systems are developed on top of C/C++/Fortran to achieve native hardware performance, and they are most often deployed on specialized hardware. 



AI/ML has taken center stage in the data engineering research community in recent years. While multidimensional array data (termed "tensors") are used for computations, AI/ML models depend on well-defined preprocessed inputs from large heterogeneously typed datasets. A good example of this is Facebook's DLRM (Deep Learning Recommendation Model) application \cite{dlrm}.

\begin{figure}[htbp]
\begin{center}
\includegraphics[width=0.35\textwidth]{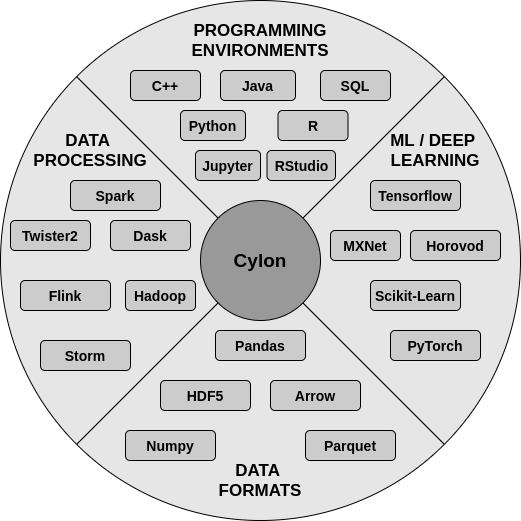}
\end{center}
\caption{High performance data engineering everywhere}
\label{fig:Cylon_port}
\end{figure}
    

\section{Cylon}\label{s:cylon} 

In designing a futuristic framework for data engineering, it is vital to pay attention to both performance and usability. An ideal data engineering framework design should be able to benefit from both Big Data and HPC worlds. With AI/ML also becoming a key driver, it is impossible for a single system to provide every feature under one roof. As an example, Apache Spark developed MLlib alongside the data analytics engine. Ultimately it lost popularity to Tensorflow and PyTorch, yet Spark is still being used as a preprocessing engine for AI/ML applications. Hence a better solution would be to create a fast and scalable framework that can universally integrate with other systems.

The goal of Cylon is to fulfill this requirement, and enable \emph{"high performance data engineering everywhere!"}\cite{widanage2020high}. We will showcase that the architecture of Cylon not only enables fast and scalable distributed data aggregations, but also provides universal integration bindings to other frameworks as shown in the Figure \ref{fig:Cylon_port}.


    \subsection{Data Model}\label{s:s:data-model}
    Cylon is a framework that mainly focuses on handling Online Analytical Processing (OLAP) workloads. Unlike Online Transaction Processing (OLTP) systems, OLAP workloads can benefit greatly from data models that have been optimized for homogeneous sequential reads. Hence Cylon has built its core on a columnar data format based on Apache Arrow\cite{apache-arrow} while providing a Table API abstraction atop a collection of data columns. 
    
    Embracing Apache Arrow's columnar format comes with many other advantages, such as inter-portability with existing popular frameworks (Spark, Numpy, Pandas, Parquet, etc.) and optimized memory operations at multiple storage levels ranging from disk (compression) to CPU cache (SIMD operations, efficient cache utilization due to contiguous data layout).
    
    \subsection{Operators}\label{s:s:operators}
    Cylon's table operators can be categorized broadly into two categories based on how they rely on the hardware:
    \begin{enumerate}
        \item Local Operators
        \item Distributed Operators
    \end{enumerate}
    
    The performance of the local operations are mainly bound by the memory (Disk, RAM and Cache) and CPU, while distributed operations are additionally  bound by the network. We currently provide the following relational and aggregation operators.
    
    \begin{table}[htb]
    \caption{Cylon Relational and Aggregation Operators}
    \begin{tabular}{|l|l|}
    \hline
    \textbf{Operator} & \textbf{Description} \\ 
    \hline
    select ($\sigma$)  & Filters out some records based on the \\
    & value of one or more columns \\
    \hline
    project ($\pi$)  & Creates a different view of the table by \\&dropping some of the columns \\
    \hline
    union ($\cup$)              & Applicable on two tables having similar \\&schemas to 
    retain all the records from \\ &both tables and remove duplicates                   \\ 
    \hline
    intersect ($\cap$)              & Applicable on two tables having similar \\& schemas to  
    retain only the records that \\& are present in both tables   \\
    \hline
    difference ($-$)              & Retains all the records of the first table, \\& while removing the matching records \\& present in the second table                   \\ 
    \hline
    join ($\bowtie$)              & Combines two tables based on the values\\& of columns.
    Cylon supports Left, Right, \\&Full, Outer and Inner \\ 
    & join modes.                   \\ 
    \hline
    Sort              & Sorts the records of the table based on a \\&specified column\\
    \hline
    Group by              & Creates multiple tables (groups) based \\&on a specified criteria\\
    \hline
    Aggregate              & Performs a calculation on a set of \\& values (records) and outputs a single value (record)\\
    \hline
    \end{tabular}
    \label{tab:operators}
    \end{table}
    
\subsection{Distributed Memory Execution}

Cylon applications can be run either in local mode or distributed mode, where local mode will be contained to a single node, and distributed mode can be scaled across a cluster of nodes. When running distributed mode, a Cylon table defined in one node can be considered a partition of a dataset that has been distributed across multiple nodes. Thus when applying most of the operators mentioned in Table \ref{tab:operators}, Cylon will be internally performing an all-to-all communication to rearrange the data partitions based on the operator's requirements. The current implementation of Cylon uses MPI at the communication layer, and is capable of using TCP, remote direct memory access (RDMA) or any other software-driven or hardware-accelerated transport layer protocol based on the availability of the resources. 
    
\section{Cylon Aggregations Architecture}\label{s:aggoperators}

Cylon aggregation operations are currently provided by the \emph{compute} API. It is broadly divided into two sections:  Aggregate operations and group by followed by aggregate operations. The following subsections will discuss the architecture and design considerations behind these operations. Figures \ref{fig:cpp_join_code} and \ref{fig:python_agg_code} shows an example code snippet of C++ and Python respectively. 

\lstset{
   basicstyle=\fontsize{7}{7}\selectfont\ttfamily
}
\begin{figure}
    \caption{Cylon Distributed Aggregations With C++}
    \begin{lstlisting}[language=C++]
int main() {
  auto mpi_config = cylon::net::MPIConfig::Make();
  auto ctx = cylon::CylonContext::InitDistributed(mpi_config);
  cylon::Status status;

  std::shared_ptr<Table> input;
  status = cylon::FromCSV(ctx, "/tmp/input.csv", input);
  CHECK_STATUS(status, "Reading csv1 failed!")

  // Sum operation
  std::shared_ptr<cylon::compute::Result> sum;
  status = cylon::compute::Sum(input, 1, output);
  CHECK_STATUS(status, "Sum failed!")

  // Group-by sum operation
  std::shared_ptr<Table> groupby;
  status = cylon::GroupBy(input, 0, {1},
                          {GroupByAggregationOp::SUM},
                          groupby);
  CHECK_STATUS(status, "Sum failed!")
  
  ctx->Finalize();
  return 0;
}
\end{lstlisting}
\label{fig:cpp_join_code}
\end{figure}

\lstset{
   basicstyle=\fontsize{7}{7}\selectfont\ttfamily
}
\begin{figure}
    \caption{Cylon Distributed Aggregations With Python}
    \begin{lstlisting}[language=Python]
from pycylon import Table, CylonContext
from pycylon.net import MPIConfig
from pycylon.io import read_csv

mpi_config = MPIConfig()
ctx = CylonContext(config=mpi_config, distributed=True)

tb = read_csv(ctx, "/tmp/input.csv")

# Sum operation
tb_sum = tb.sum(1)

# Group-by sum operation
tb_gby_sum = tb.groupby(0, 1, [AggregationOp.SUM])

ctx.finalize()
\end{lstlisting}
\label{fig:python_agg_code}
\end{figure}

    \subsection{Aggregate Operations}\label{s:agg}

    As explained in Section \ref{s:cylon}, Cylon employs Apache Arrow \cite{apache-arrow} columnar data structure underneath it. A table is partitioned into multiple shards across distributed processes. A column of a table may contain multiple chunks of data. We define an aggregation operation as "a reduction of all values in a (distributed) column". Based on this setup, we identify core components to implement an aggregation operation on columnar data tables. 
    \begin{itemize}
        \item Intermediate and final result definition 
        \item Bulk Reduction 
        \item Element-wise Reduction 
        \item Communication of intermediate results 
        \item Final result conversion 
    \end{itemize}
    
\begin{figure}[]
\begin{center}
\includegraphics[width=0.5\textwidth]{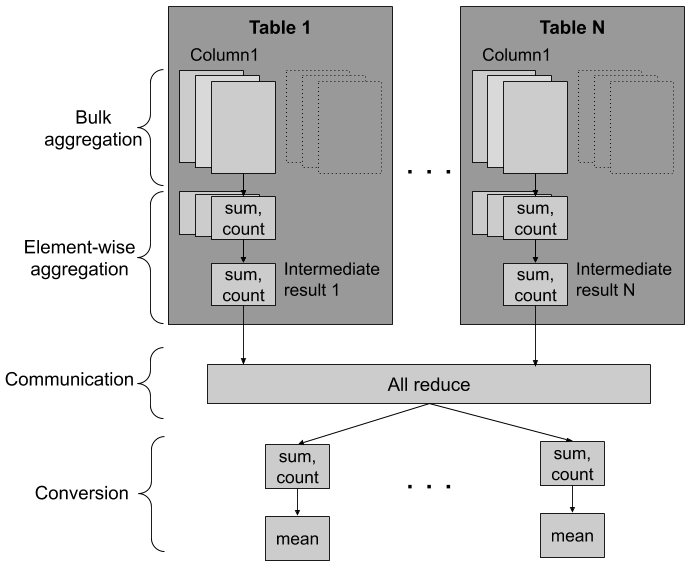}
\end{center}
\caption{Operation flow of \emph{mean} aggregation}
\label{fig:agg}
\end{figure}

    This approach has been widely adopted in the OLAP columnar database domain (ex: ClickHouse Database \cite{clickhouse-db}). Additionally, a similar approach has inspired the Apache Arrow \emph{Chunked Array} aggregation operations in their \emph{Table API}. Figure \ref{fig:agg} depicts the operation flow of a mean aggregation. 
        
        \subsubsection{Intermediate and final result definition}
        We posit that distributed aggregation operations can be more complex than a trivial \emph{MPI{\textunderscore}Allreduce} operation. Operations would need to track multiple intermediate values before arriving at the global final result. It all depends on the "moment" of the statistic/operation. For example, to compute the \emph{mean} (first moment statistic), the sum and count would have to be tracked individually for every column chunk in each shard of the table. For higher moments, such as \emph{standard deviation}, sum of squares, sum, and count would need to tracked. These intermediate results/state would then need to be converted to the final result. An additional consideration would be the data types of intermediate results. Hence an aggregation operation would have to define these attributes.
        
        \subsubsection{Bulk Reduction}
        The role of bulk reduction is to aggregate partial array data into an intermediate result/state. Columnar data being available on contiguous memory locations allows fast and efficient bulk reductions. These operations benefit from efficient CPU cache and registry usages and are further optimized from SIMD (Single-Instruction-Multiple-Data) instructions. Thus it is an obvious choice to enable bulk aggregation capability, as it will be used to reduce partitioned data into individual elements of intermediate results. 
        
        Bulk aggregations become sub-optimal if the data elements are randomly distributed in the column. A good example of such a situation is group by aggregations. These will be discussed in more detail in Section \ref{s:group}. 
        
        Cylon uses Apache Arrow \cite{apache-arrow} Compute API for bulk aggregations on chunked arrays. Arrow Compute API supports optimized aggregation kernels for 
        
        \subsubsection{Element-wise Reduction}
        Element-wise aggregations would need to be used when aggregating multiple intermediate results/states. These also form the basic aggregation function. In a scattered data environment/row-based data distribution, element-wise data reduction could be more dominant, as we see in Twister2 Keyed-Reduce operations \cite{twister2}. 
        
        \subsubsection{Communication of intermediate results}
        Since the intermediate results are distributed across processes, it would require additional communication operations to arrive at the final result. These communication operations could be semantically similar to reduce/all-reduce. This would aggregate intermediate results element-wise. Cylon currently uses OpenMPI \cite{open_mpi} as the communication fabric and reduces each element of the intermediate results individually. Another approach would be to use a custom MPI\textunderscore Op that corresponds to the aforementioned element-wise aggregations.
    
        \subsubsection{Final result conversion}
        At the point of returning the aggregated function, the final result would be calculated.

    \subsection{Group By Operations}\label{s:group}
    
    Group by is a widely used operation in traditional Big Data analysis. There are multiple approaches for group by execution, but the main idea is to bring rows together based on a particular key/index column (one or more) and apply an aggregation operation on the rows with the same key. This is semantically equivalent to the \emph{Map-Reduce} \cite{dean2008mapreduce} Big Data execution paradigm. \emph{Pivoting} is another derivation of group by operation that has become very popular on tabular data.

\begin{figure}[]
\begin{center}
\includegraphics[width=0.5\textwidth]{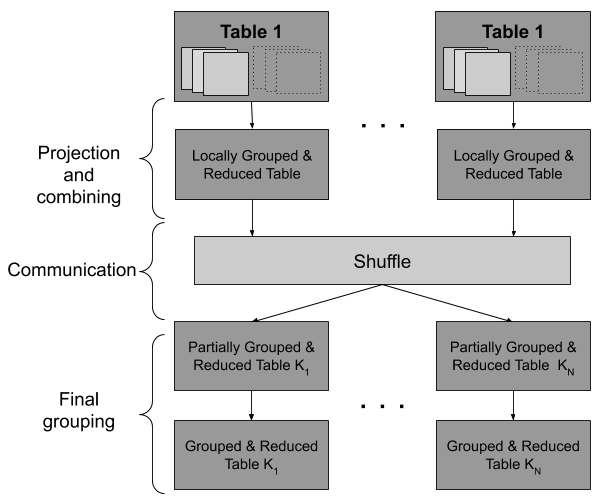}
\end{center}
\caption{Operation flow of Cylon group by operation}
\label{fig:grp}
\end{figure}

    Cylon currently supports two approaches for group by execution: Hash-based and Pipeline-based. Figure \ref{fig:grp} depicts the typical execution flow of group-by operation. 
    
        \subsubsection{Early Aggregation}
        
        \emph{Early aggregation} is a common technique used in traditional RDBMS databases that could speed up group by aggregation operations \cite{larson1997grouping}. In a distributed processing environment, early aggregation helps in reducing the communication overhead and memory required for intermediate results. This early aggregation is semantically equal to the element-wise aggregation step from Section \ref{s:agg}, but would be operating on multiple rows. Cylon takes advantage of this technique for its hash group by implementation, which is explained in the following section.

        \subsubsection{Indices of Groups}\label{s:approaches}
        A common approach being used by frameworks such as Python Pandas \cite{pandas}, Apache Spark \cite{apache-spark}, etc., is treating group by as a separate operation which returns the indices corresponding to each group. The aggregation function will subsequently be applied for these groups. Similar to \emph{Join} operations, indices of groups can be generated by either using \emph{hash-based} or \emph{sort-based} approach \cite{graefe1993query}. This effectively allows multiple aggregation operations on the same groups. Even though this seems like a very intuitive approach, it could have a performance penalty, especially for columnar data. Since values will be aggregated for each group, value columns would be accessed randomly, which could lead to poor cache performance. 
        
        Furthermore, the concept of indices of groups becomes rather obscure in a distributed table setup. In this case, groups could well be partitioned across table shards. Tables would have to be shuffled to bring all the indices to the same process. 

        Sorting can be further extended to \emph{sorting the entire table}. This could benefit columnar data aggregations because the grouped data will be on consecutive memory locations. Then the aggregation kernels could call the \emph{bulk aggregation} on value column slices. This would be an efficient execution, provided that there is a sizeable number of records in each group, but as discussed by Muller et al \cite{muller2012depth}, this leads to higher maintenance costs which could hinder overall performance (Refer Section \ref{s:pipeline}). 
        
        \subsubsection{Group by with early aggregations}
        In traditional SQL, group by queries are always accompanied by aggregation operations. As mentioned in the previous section, this allows early aggregations even while determining groups. 
        
        Cylon currently supports this approach with hash-based grouping. While creating the hash table, the values will be aggregated into intermediate values and later be written to the locally grouped table. Since Cylon works on a distributed environment, these local results would have to be shuffled amongst the processes. Then the resultant table will be grouped again to aggregate the intermediate results of the same groups. 
        
        \subsubsection{Using Local Combiner} \label{s:local_combiner}

        The worst-case scenario of the above-mentioned approach can be asymptotically analyzed as follows. Let us take a distributed table partitioned across $P$ processes, with a total $N$ records in $G$ unique groups ($ G \leq N $).  Then for \emph{each process}, 
        \begin{enumerate}
            \item Local group by (Combiner) $ = O(N/P)$
            \item Shuffle communication $= O(G)$ or $O(N/P)$ without combiner
            \item Final local group by $= O(G)$ or $O(N/P)$ without combiner
        \end{enumerate}        
        From this analysis, it is evident that the ratio $N/P : G$ is very important. If these values are relatively similar (i.e. there is less duplication amongst keys), then steps 1 and 3 would take comparable amounts of time, leading to increased total time for operation. In such a scenario, dropping the combiner step could improve the total execution time. Experimental results to support this scenario are provided in Section \ref{s:combiner}. 
        
        \subsubsection{Pipeline Group by} \label{s:pipeline}
        Pipeline Group by is a special case of group by operations that can be applied on \emph{sorted} tables. In the distributed setup, it is sufficient to have the table shards sorted locally. The term \emph{Pipeline Group By} seems to have originated from the Vertica \cite{vertica} columnar database. As explained in Section \ref{s:approaches}, this could make use of bulk aggregation operations provided that there is sufficient key duplication ($ G <<< N/P $). Nevertheless, this approach significantly reduces the memory footprint of the group by operation, and allows further optimizations using multi-threading. 
        
        Cylon also supports this approach. The experimental results are provided in Section \ref{s:hashpipeline}.

\section{Experiments}\label{s:experiments}

We analyzed the strong scaling performance of Cylon for the following scenarios and compared the performance against popular Big Data analytics framework Apache Spark \cite{apache-spark}. Furthermore we have analyzed the performance of Cylon's aggregation implementation on the following aspects. 

\begin{enumerate}
    \item Strong scaling performance comparison between Cylon vs. Spark on aggregates and group by operations. 
    \item The effect of group size on the local combiner step
    \item Hash group by vs. pipeline group by
    \item Overhead comparison between Cylon's Python and Java bindings. 
\end{enumerate}

    \subsection{Setup}
    The tests were carried out in a cluster with 10 nodes. Each node is equipped with Intel\textsuperscript{\textregistered} Xeon\textsuperscript{\textregistered} Platinum 8160 processors. A node has a total RAM of 255GB and mounted SSDs were used for data loading. Nodes are connected via Infiniband with 40Gbps bandwidth. 
    
    \emph{Software Setup}: Cylon was built using g++ (GCC) 8.2.0 with OpenMPI 4.0.3 as the distributed runtime. \emph{Mpirun} was mapped by nodes and bound sockets. Infiniband was enabled for MPI. For each experiment, a maximum of 16 cores from each node were used, reaching a maximum parallelism of 160.
    
    Apache Spark 2.4.6 (hadoop2.7) pre-built binary was chosen for this experiment alongside its PySpark release. Apache Hadoop/HDFS  2.10.0 acted as the distributed file system for Spark, with all data nodes mounted on SSDs. Both Hadoop and Spark clusters shared the same 10-node cluster. To match MPI setup, \emph{SPARK\_WORKER\_CORES} was set to 16 and \emph{spark.executor.cores} was set to 1. Additionally we also tested PySpark with \emph{spark.sql.execution.arrow.pyspark.enabled} option, which would allow PyArrow underneath PySpark dataframes. 
    
    This notation will be used in the following sections. 
    \begin{itemize}
        \item $N$, Total number of rows in the \emph{distributed} table
        \item $P$, Parallelism/number of partitions 
        \item $N_P$, Rows per partition
        \item $G$, Number of unique groups in the dataset ($ G \leq N $)
    \end{itemize}

    \emph{Dataset Formats}: For strong scaling test cases, CSV files were generated with two columns (an int\_64 as index and a double as value). The same files were then uploaded to HDFS for the Spark setup and output counts were checked against each other to verify the accuracy. Times were recorded only for the corresponding operation (no data loading time considered). All numbers are generated using a uniform random distribution. 
    
    \subsection{Scalability}

\begin{figure*}[htb]
\centering
\begin{tabular}[width=\textwidth]{|c|c|}
\hline
 Sum & Group-By Sum
\\
\hline
\subf{\includegraphics[width=80mm]{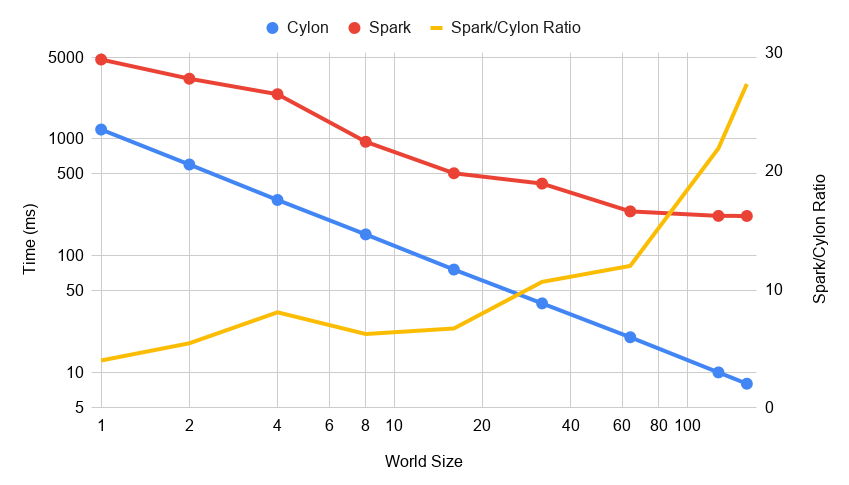}}{(a)}
&
\subf{\includegraphics[width=80mm]{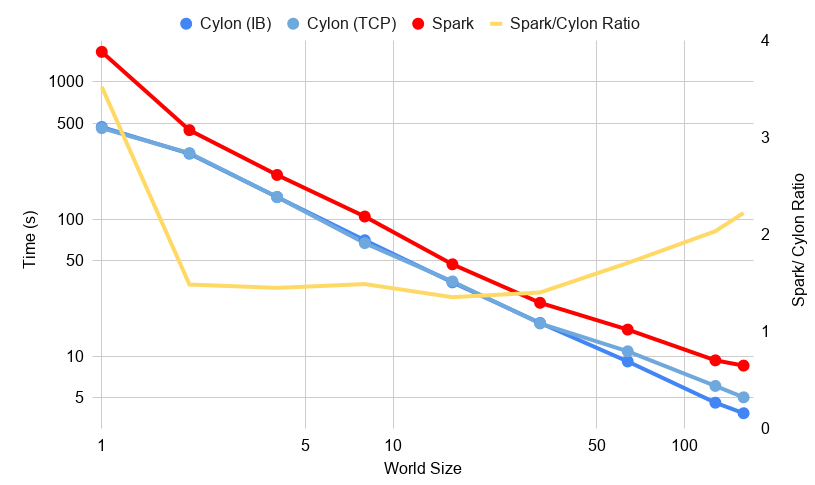}}{(b)}
\\
\hline
\end{tabular}
\caption{Cylon vs. Spark strong scaling (Log-Log plots) for $N = 1 \times 10^9$, $G=0.99\times 10^9$ }
\label{fig:cylon-compare}
\end{figure*}
    
    To test the scalability of the aggregation operations, we varied the parallelism from 1 to 160 while keeping total work at 200 million rows per table. The results for both aggregation and group by operations are shown in Figure \ref{fig:cylon-compare}. We have chosen 1 billion records ($N = 1 \times 10^9$) and a $G = 0.99 \times 10^9$, which produces $1.01$ average rows per group. This case reflects the worst-case scenario for group by operations (it has no effect on aggregations). 
    
    \subsubsection{Aggregation Scaling}

    We have tested the performance of calculating the sum of a column. The results are shown in  Figure \ref{fig:cylon-compare}(a). As evident from the graphs, Cylon shows almost perfect linear scaling as expected. In comparison, Spark scales much slower. The speed-up of Cylon over Spark increases from \emph{4x} to \emph{27x} as the number of processes increases. 
    
     \subsubsection{Group By Scaling}
     
    For group by operations, we have calculated the time spent on group by followed by sum. Group by operation results are shown in  Figure \ref{fig:cylon-compare}(b). Cylon seems to demonstrate linear strong scaling. This is expected as the execution becomes increasingly communication-dominant for higher values of $P$. These results coincide with the Cylon C++ performance in our prior publication \cite{widanage2020high}. In contrast, Spark seems to plateau earlier than Cylon, leaving a maximum speedup around 2x at 160 processes.

    \subsection{Local Combiner vs. Group Size Group By's}\label{s:combiner}

    To assess the effect of group size on the overall performance, we have varied the average rows per group from $(0.99)^{-1} \approx 1.01$ to $10,000$ while keeping $N$=200 million. This will effectively change $G$ from $198\times 10^6$ to $20\times 10^3$. We have considered two cases: $P=64$ and $P=128$. 
    
    As the number of average rows per group increases, the total time reduces. With the local combiner, this effect is much more prominent. The speed-up with the combiner changes from 0.5x to 3-4x. This corresponds to the analysis given in Section \ref{s:local_combiner}. When $G=20 \times 10^3$, $N_P = 1.25 \times 10^6$, therefore having a local combiner significantly reduces the downstream workloads.
    
\begin{figure}[]
\begin{center}
\includegraphics[width=0.47\textwidth]{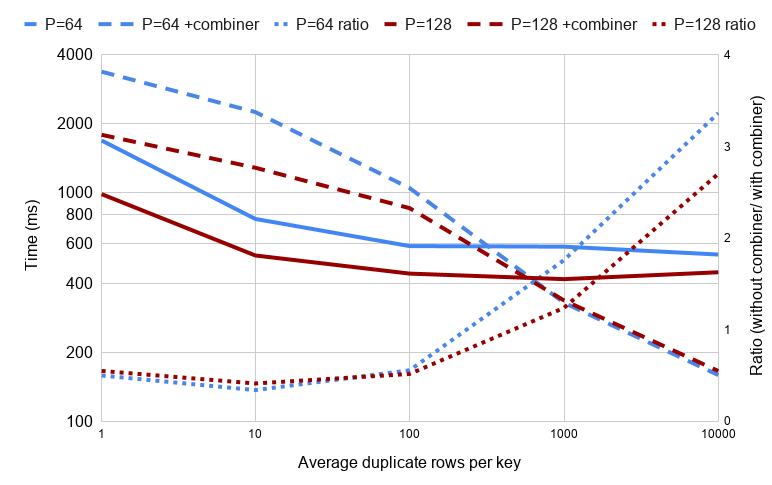}
\end{center}
\caption{Local combiner step vs. key distribution for group-by sum (Log-log plot)  $N=2\times10^6$}
\label{fig:dup}
\end{figure}
    
    \subsection{Hash Group By vs. Pipeline Group By} \label{s:hashpipeline}
    
\begin{figure}[]
\begin{center}
\includegraphics[width=0.5\textwidth]{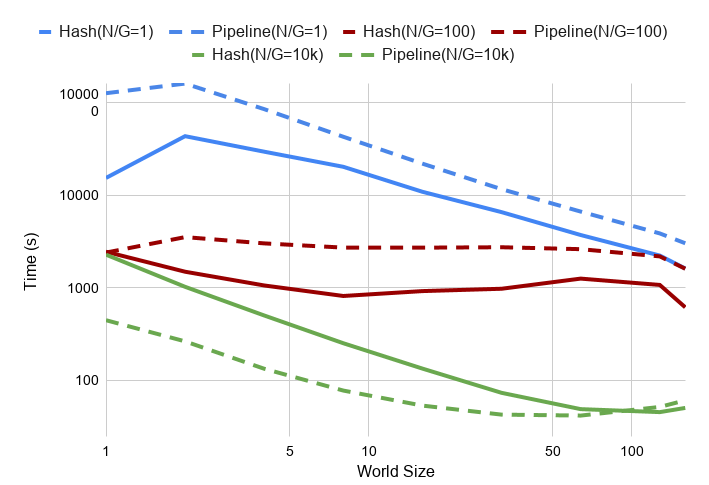}
\end{center}
\caption{Hash group-by [solid lines]  vs pipeline group-by [dashed lines] followed by sum (Log-log plot)   $N=2\times10^6$}
\label{fig:hash_pip}
\end{figure}

    Figure \ref{fig:hash_pip} depicts hash-based group by vs. pipeline group by for a \emph{locally sorted} distributed table with $N=$ 200 million records. Tests were carried out varying the average number of rows per group ($N/G$) for 1, 100, and 10,000. From the  graphs, it is evident that pipeline group by performance is adversely affected by the lack of work for bulk aggregations in a group slice. As the number of rows per group increases, pipeline group by becomes more effective than the hash group by. 
    
    \subsection{Switching between C++, Python \& Java}

    All of the previous experiments were done on various aspects of the aggregation performance. Since Cylon is engineered to be an integrating approach for all data engineering applications, it is worthwhile evaluating the overheads while switching between language bindings. Table \ref{t2} shows the time ratio for Inner-Join (Sort) for 200 million rows while changing the number of workers. It is clear that the overheads between Cylon and its Cython Python bindings and JNI Java bindings are negligible.
    
\begin{table}[htb]
\caption{Cylon, PyCylon vs. JCylon}
\begin{center}

\begin{tabular}{|c|c|c|}
\hline
\textbf{World size} & \textbf{PyCylon/Cylon} & \textbf{JCylon/Cylon} \\ \hline
16                  & 1.00                   & 1.07                  \\ \hline
32                  & 0.99                   & 1.04                  \\ \hline
64                  & 1.01                   & 1.04                  \\ \hline
128                 & 1.00                   & 1.01                  \\ \hline
\end{tabular}
\end{center}
\label{t2}
\end{table}
    

\section{Related Work}\label{s:related-work}

Relational databases, Structured Query Language (SQL), and disk-based analytics frameworks such as Apache Hadoop \cite{apache-hadoop} are at the heart of traditional Big Data aggregations. Considered the first generation of Big Data analytics, Hadoop revolutionized the industry by introducing the MapReduce programming model \cite{dean2008mapreduce}. Apache Spark \cite{apache-spark} and Apache Flink \cite{flink2015} subsequently overtook Hadoop by providing faster user-friendly APIs. These also benefited from the boost in hardware advancements that allowed them to perform Big Data processing in-memory.

More recently, Python Pandas Dataframes \cite{pandas} emerged as the preferred data analytics abstraction amongst the data science and engineering community. Even though Pandas are limited in performance and scalability, they provide an extremely convenient programming environment for data processing. Programmer usability became so important that frameworks such as Spark and Flink provided Python wrappers around their data abstractions. But with Java and Python runtimes not being inherently compatible with each other, this hindered their performance. Dask Distributed \cite{dask} is a distributed DataFrame abstraction on Python Pandas, and Modin \cite{modin}\cite{petersohn2020towards} generalized the Pandas API. Later, CuDF \cite{cudf} emerged as a DataFrame abstraction that could be used for ETL pipelines on top of GPU hardware. 

Aggregations have been a widely studied area in the database domain for decades. Smith et al \cite{smith1977database} formally defined the fundamentals of aggregation operations. Later Gray et al \cite{gray1997data} comprehensively analyzed aggregation functions and also introduced a \emph{Data Cube operator} extending the usual group by aggregation behavior. Larson et al \cite{larson1997grouping} studied impact on grouping by early aggregation. More recently, greater focus has been given to columnar datastores such as Vertica (commercialized version of C-Store) \cite{vertica}, MonetDB, GreenPlum, ClickHouse \cite{clickhouse-db}, etc., for their efficient access patterns in OLAP query processing. Abadi et al \cite{abadi2013design} discussed the general design and implementation of such databases. 

\section{Conclusion}\label{s:conclusion}

Big Data analytics and data engineering have experienced an exponential growth in both research effort and applications. This has expanded the boundaries of traditional stand-alone data analytics solutions (databases, analytics frameworks, etc.) beyond their capabilities. But no framework on its own can fulfill all these requirements. Hence we believe that there is an opportunity for a fast, flexible and integrating framework that could bring all these environments together. Cylon strives to serve this purpose. 

Cylon's C++ core allows efficient data analytics implementations, and in this paper we confirmed that data aggregations can also benefit from the same architecture. It allows data aggregations to take advantage of both Big Data and high performance computing domains. Another qualitative requirement of data engineering is to write ETL pipelines in popular languages like Java and Python without compromising on performance. Offering Cython-based Python APIs for compute kernels means less overhead across the runtimes and good scaling\cite{pycylon-hpc}.  

From our experiments, we can confirm that Cylon's architecture achieves superior performance and scalability than the state-of-the-art Big Data systems, and universally integrates with cross-platform frameworks. It also shows potential for further improvements.

\section{Future Work}\label{s:future-work}

Cylon is a project still in its early stages, and we believe that there is a substantial potential for more performance and usability improvements. The current compute kernels do not take into account factors such as NUMA boundaries, in-cache performance, etc. As the number of processes inside a node increases, we can expect resource contention for memory bandwidth and L1/L2 caches. Polychroniou et al \cite{polychroniou2014comprehensive} show that these factors play a vital role in sorting and hashing operations. Furthermore, we believe that the relational algebraic operations such as joins, groupings, etc. can benefit from efficient in-place sorting operations.

Currently, Cylon aggregates, group by operations, and their corresponding communication APIs are rather disjointed, and we believe that these APIs can be combined into a more uniform distributed computing API. We are also evaluating the possibility of using UCX \cite{ucx} as another communication fabric. While developing Twister2 \cite{twister2}, we have experienced that UCX is an effective communication abstraction for distributed systems. Additionally we believe compute kernels should be able to make use of specialized computational subsystems such as GPUs and CUDA compute routines, FPGAs, etc., and we believe that Cylon's  architecture supports such integration.

We agree with Petersohn et al's \cite{petersohn2020towards} suggestion that conforming to the Pandas dataframe API is an important feature for data engineering tools. We are currently developing a dataframe API based on Modin, and as such Cylon would be another distributed back-end for Modin. To expand our compute kernels, we are currently focusing on supporting distributed computing on array data structures. In supporting diverse data formats, we will be integrating HDF5, Parquet, GPFS/Lustre data loading and data processing in a future software release.

\section*{Acknowledgments}
This work is partially supported by the National Science Foundation (NSF) through awards CIF21 DIBBS 1443054, nanoBIO 1720625, CINES 1835598 and Global Pervasive Computational Epidemiology 1918626. We thank the FutureSystems team for their support with the Juliet and Victor infrastructure.

\balance
\bibliographystyle{bibliography/IEEEtran}

\bibliography{bibliography/IEEEexample}

\end{document}